\documentstyle[aps,multicol]{revtex}
\def\be{\begin{eqnarray}}
\def\ee{\end{eqnarray}}

\begin{document}
\title{Quantum Secret Sharing}
\author{
Mark Hillery$^{1}$,
Vladim\'{\i}r Bu\v{z}ek$^{2}$, and Andr\'{e} Berthiaume$^{3}$
}
\address{
$^{1}$Department
of Physics \& Astronomy, Hunter College, CUNY,
695 Park Avenue, New York, NY 10021, USA
\newline
$^{2}$
Institute of Physics, Slovak Academy of Sciences, D\'{u}bravsk\'{a} cesta 9,
842 28 Bratislava, Slovakia
\newline
$^{3}$
 DePaul University, School of CTI,
243 S. Wabash Ave, Chicago, Il 60604-2302, USA
}

\date{1 December, 1998}
\maketitle

\begin{abstract}
Secret sharing is a procedure for splitting a message
into several parts so that no subset of parts is
sufficient to read the message, but the entire set
is. We show how this procedure can be implemented
using GHZ states.  In the quantum case the presence
of an eavesdropper will introduce errors so that
his presence can be detected.  We also show how
GHZ states can be used to split quantum information
into two parts so that both parts are necessary to
reconstruct the original qubit.

\end{abstract}

\begin{multicols}{2}

\section{Introduction}
Suppose Alice, who is in New York, wants to have an action 
taken on her behalf in Prague.  There she has two agents, 
Bob and Charlie, who can carry it out for her, but she 
knows that one of them, and only one, is dishonest, and 
she does not know which is the honest one.  She cannot
simply send a message to both, because the dishonest one
will try to sabotage the action, but she knows that if 
the two of them carry out it together, the honest one 
will keep the dishonest one from doing any damage.  What
can she do?

Classical cryptography provides an answer which is known 
as secret sharing \cite{crypto}.  It can be used, for
example, to guarantee that no single person can open 
a vault, has access to an industrial secret, or can 
launch a missile with a nuclear warhead, but two together 
can.  This means that for security to be breached, two 
people must act in concert, thereby making it more 
difficult for any single person who wants to gain illegal 
access to the secret information; he must convince
the other party to go along, and he risks discovery in 
the process.

How can Alice implement this procedure? From her original 
message, she creates two coded messages one of which 
is sent to Bob and the other to Charlie.  Each of the 
encrypted messages contains no information about her 
original message, but together they contain the complete 
message.  Therefore, neither Bob nor Charlie alone can 
find out what Alice wants to do, but the two of them 
acting together can.  This can be accomplished by 
taking the original message, which we can think of 
as a binary bit string, and adding to it a random 
bit string of the same length.  The addition is done 
modulo 2 and bitwise.  
Alice then takes this string and a copy of the random 
string and sends one to Bob and the other to Charlie.  
At this point neither is in a position to learn Alice's 
message.  However, if they get together and add their 
two strings together, bitwise and modulo 2, Alice's
message emerges.  There are also classical protocols
which allow Alice to split her message into more than
two parts.

So far we have not mentioned the problem of eavesdropping,
but this is something Alice must consider.  If either a
fourth party or the dishonest member of the Bob-Charlie
pair gains access to both of Alice's transmissions, then 
they can learn the contents of her message.  
Eavesdroppers can, however, be defeated by using quantum
cryptographic protocols.  Quantum cryptography provides 
for the secure transmission of information by
enabling one to determine whether an eavesdropper 
has attempted to gain information about the key
which is being used to encode the message 
\cite{qcrypt1,qcrypt2,qcrypt3}.  
If not, the key can be used and the information sent by 
using it will be secure, and if an eavesdropper has
been detected, then one has to establish a new key.

We would like to show that it is possible to combine
quantum cryptography with secret sharing in
a way that will allow one
to determine whether an eavesdropper has been active
during the secret sharing protocol.  The most obvious
way of doing this is simply for Alice to use quantum
cryptographic protocols to send each of the bit 
strings which result from the classical secret sharing 
procedure, and this method will work.  It is, however, 
awkward.  One first must establish mutual keys among 
different pairs of parties, in this case one for Alice 
and Bob, and another for Alice and Charlie, and then 
implement the classical procedure.  The classical 
procedure, it should be pointed out, becomes more and 
more complicated the larger the number of pieces into 
which one wants to split the message.  We would like to 
explore an alternative which uses quantum mechanics to 
do both the information splitting and the eavesdropper 
protection simultaneously.  By using multiparticle
entanglement, it eliminates the need to perform
the classical secret-splitting procedure altogether.

The method for splitting a message into two parts, 
which we present here uses maximally 
entangled three-particle states, or GHZ states, and it
can be easily extended in two different ways.  First, 
it can be modified to allow Alice to send
a string of qubits to Bob and Charlie in such a way 
that only by working together can they determine what
the string is.  In this case it is quantum information 
which has been split into two pieces, neither of which
separately contains the original information, but whose
combination does.  Second, the procedure can also be 
generalized to more than three parties, and we show
explicitly how it works with four.

GHZ states have already found a number of uses.  They 
form the basis of a very stringent test of local 
realistic theories \cite{GHZ}.  Recently it was also
proposed that they can used for cryptographic
conferencing or for multiparticle generalizations of
superdense coding \cite{Bose}.  In addition, related
states can be used to reduce communication complexity
\cite{Cleve}.  Quantum secret sharing represents yet
another application.

\section{GHZ States and Secret Sharing}
Let us suppose that  Alice, Bob, and Charlie each have  
one particle from a GHZ triplet which is in the state 
\begin{eqnarray}
\label{1} 
|\psi\rangle =\frac{1}{\sqrt{2}}(|000\rangle + |111\rangle ).
\end{eqnarray}
They each choose at random  whether 
to measure their particle in the $x$ or $y$ direction.  
They then announce publicly in which direction they 
have made a measurement, but not the results of their 
measurements. Half the time, Bob and Charlie, by 
combining the results of their measurements, can 
determine what the result of Alice's measurement was.
This allows Alice to establish a joint key with Bob
and Charlie, which she can then use to send her message. 

Let us see how this works in more detail.  Define 
the $x$ and $y$ eigenstates
\begin{eqnarray}
|+x\rangle = \frac{1}{\sqrt{2}}(|0\rangle + |1\rangle );&
\hspace{0.5cm}&|+y\rangle = \frac{1}{\sqrt{2}}(|0\rangle
+i|1\rangle ); \nonumber \\
|-x\rangle = \frac{1}{\sqrt{2}}(|0\rangle - |1\rangle );&
\hspace{0.5cm}&|-y\rangle = \frac{1}{\sqrt{2}}(|0\rangle
-i|1\rangle ).
\label{2}
\end{eqnarray}
We can see the effects of measurements by Alice and Bob
on the state of Charlie's particle, if 
we express the GHZ state in different ways.  Noting that
\begin{equation}
|0\rangle = \frac{1}{\sqrt{2}}(|+x\rangle + |-x\rangle );
\hspace{0.1cm}
|1\rangle = \frac{1}{\sqrt{2}}(|+x\rangle - |-x\rangle )
\label{3}
\end{equation}
we can write
\begin{eqnarray}
|\psi\rangle &= & \frac{1}{2\sqrt{2}}\left[(|+x\rangle_{a}
|+x\rangle_{b}+|-x\rangle_{a}|-x\rangle_{b})
(|0\rangle_{c} + |1\rangle_{c}) \nonumber \right.\\
 & &+\left.(|+x\rangle_{a}|-x\rangle_{b}+|-x\rangle_{a}
|+x\rangle_{b})(|0\rangle_{c}-|1\rangle_{c})\right].
\label{4}
\end{eqnarray}
This decomposition of $|\psi\rangle$ tells us what
happens if both Alice and Bob make measurements in
the $x$ direction.  If they both get the same result,
then Charlie will have the state $(|0\rangle_{c}
+|1\rangle_{c})/\sqrt{2}$, and if they get different
results he will have the state $(|0\rangle_{c}
-|1\rangle_{c})/\sqrt{2}$.  He can determine which
of these states he has by performing a measurement 
along the x direction.  The following table summarizes
the effects of Alice's and Bob's measurements on
Charlie's state:
\begin{center}
\begin{tabular}{c|c|c|c|c|c|} 
\multicolumn{6}{c}{Alice}  \\ \cline{2-6}
 & & $+x$ & $-x$ & $+y$ & $-y$ \\ \cline{2-6}
  &$+x$ & $|0\rangle + |1\rangle$ & $|0\rangle - |1\rangle$
 &$|0\rangle -i |1\rangle$ & $|0\rangle + i|1\rangle$ \\
 \cline{2-6}
 Bob
  &$-x$ & $|0\rangle - |1\rangle$ & $|0\rangle + |1\rangle$ 
 &$|0\rangle + i|1\rangle$ & $|0\rangle -i |1\rangle$ \\
 \cline{2-6}
  &$+y$ & $|0\rangle -i |1\rangle$ & $|0\rangle + i|1\rangle$
 & $|0\rangle - |1\rangle$ & $|0\rangle + |1\rangle$ \\
 \cline{2-6}
  &$-y$ & $|0\rangle + i|1\rangle$ & $|0\rangle -i |1\rangle$
 & $|0\rangle + |1\rangle$ & $|0\rangle - |1\rangle$ \\
 \cline{2-6}
\end{tabular}
\end{center}
Alice's measurements are given in the columns and Bob's 
are given in the rows.  Charlie's state, up to 
normalization, appears in the boxes.  From the table it
is clear that if Charlie knows what measurements Alice
and Bob made (that is, $x$ or $y$), he can determine
whether their results are the same or opposite, and
also that he will gain no knowledge of what their results
actually are.  Similarly, Bob will not be able to 
determine what Alices's result is without Charlie's
assistance, because he does not know if his result is
the same as Alice's, or the opposite of hers.

With each party choosing to make $x$ or $y$ measurements
at random, only half of the GHZ triplets will give
useful results.  For example, if Alice and Bob both
measure their particles in the $x$ direction, Charlie
must also measure his in the $x$ direction in order to
determine whether the results of Alice's and Bob's 
measurements are correlated or anticorrelated; if
he measures in the $y$ direction he gains no
information. Because Charlie is choosing his 
measurement direction at random, he will only choose
correctly half the time.  This is why all three 
parties must announce the directions of their
measurements, so that they can decide whether to keep 
or to discard the results from a given triplet.  This
announcement should be done in the following way:  
Bob and Charlie both send to Alice the direction of
their measurements who then sends all three 
measurement directions to Bob and Charlie.  

Before presenting a more general discussion of 
eavesdropping, we shall consider a specific 
situation in order to show that it can be detected.
Suppose that Bob is dishonest and that he has
managed to get a hold of Charlie's particle as
well as his own.  He then measures the two
particles and sends one of them on to Charlie.
His object is to discover what
Alice's bit is, without any assistance from Charlie,
and to do so in a way that cannot be detected.
Alice has measured her particle
in either the $x$ or $y$ direction, but Bob does
not know which.  He would like to measure the quantum
state of his two-particle system, but because he
does not know what measurement Alice made, he does
not know whether to make his in the $(|00\rangle\pm
|11\rangle )/\sqrt{2}$ basis or in the 
$|00\rangle\pm i|11\rangle )/\sqrt{2}$ basis.  
Choosing at random he has a
probability of $1/2$ of making a mistake.  If
he chooses correctly, he will know, for valid
combinations of measurement axes, what the result
of Charlie's measurement is from the result of his
own, and this means he will then know what Alice's
bit is.  For example, if Alice measured in the $x$
direction and found $|+x\rangle$, then the state
Bob receives is $|00\rangle +|11\rangle )/\sqrt{2}$.
If Bob now measures in the $|00\rangle\pm 
|11\rangle )/\sqrt{2}$ basis, he knows what the two-
particle state is, and because
\begin{equation}
\frac{1}{\sqrt{2}}(|00\rangle +|11\rangle ) =\frac{1}
{\sqrt{2}}(|+x\rangle |+x\rangle +|-x\rangle |-x\rangle).
\label{5}
\end{equation}
Bob knows that Charlie's measurement will produce a
result identical to his.

What happens if he is wrong?  Suppose that Alice has
measured her particle in the $y$ direction and
that Bob measures his particles in the $(|00\rangle
\pm |11\rangle )/\sqrt{2}$ basis.  He has a 
probability of $1/2$ of getting either basis vector.
He now sends one of his particles to Charlie, and
both Bob and Charlie measure their particles.
Because Alice measured $y$, in order for this
round of measurements to produce a valid key bit,
Bob and Charlie must make different measurements,
i.\ e.\ one must measure $x$ and the other $y$.
We note that in the $(|00\rangle\pm |11\rangle )/
\sqrt{2}$ basis there is no correlation between
$x$ and $y$ measurements, for example
\begin {eqnarray}
\frac{1}{\sqrt{2}}(|00\rangle &+& |11\rangle ) = 
\frac{1}{2}\left[e^{-i\pi /4}
(|+x\rangle |+y\rangle +
|-x\rangle |-y\rangle)
\right.  \nonumber \\
 &+&\left.
e^{i\pi /4}(|+x\rangle |-y\rangle +|-x\rangle |+y\rangle) 
\right] .
\label{6}
\end{eqnarray}
Therefore, in half the situations the results of the
measurements will be wrong.  If, for example, Alice
found $|+y\rangle$ and Bob found $|+x\rangle$, then
Charlie should measure $|-y\rangle$ if he measures
his particle in the $y$ direction, but because of 
Bob's measurement, he has a probability of $1/2$
of finding $|+y\rangle$.  The overall probability
of an error in this cheating scheme is $1/4$, one
half of picking the wrong basis and then one
half of getting the wrong result.

There are two additional points to notice here.  First,
if Bob were able to learn the direction of Alice's
and Charlie's measurements before having to reveal 
his, he could cheat more successfully.  In the cases 
in which he made the wrong measurement, Bob could
simply tell Alice a measurement direction which would
cause the results from that triplet to be thrown 
out.  Alice and Charlie would, however, notice a
higher than usual failure rate, 75 \% as opposed to 
50 \%, which would tell them
that something unusual was happening.  Insisting that
Bob send a measurement direction to Alice before
learning what kind of measurement Alice and Charlie 
makes this kind of cheating more difficult.  
Second, there is 
also the possibility that Bob could lie at certain points
in the procedure; he could lie about his measurement
direction or about the result of his measurement.  In 
the cheating scheme considered above he
gains, however, nothing by doing so. 

Now let us look at a more general situation.  We
assume that there is an
eavesdropper, Eve (who could also be either Bob 
or Charlie).  Her problem, as in the example which
we just discussed, is that she
does not know what bases have been or will be used
to measure the particles.  If she measures them
herself, and chooses the wrong bases, she will
introduce errors which Alice, Bob and Charlie
will be able to detect by publicly comparing a subset
of their measurements.

In order to show this for a large class of
measurements, let us
assume that Eve has been able to entangle an ancilla
with the three particle state which Alice, Bob, and
Charlie are using.  At some later time she can measure
the ancilla to gain information about the measurement
results of Alice, Bob and Charlie.  The state
describing the state of the three particles and the
ancilla is
\begin{equation}
|\Psi\rangle = \sum_{j,k,n=0}^{1}|jkn\rangle_{3}
|R_{jkn}\rangle_{\xi} ,
\label{7}
\end{equation}
where $|jkn\rangle_{3}$ is a state of the three particles,
and $|R_{jkn}\rangle_{\xi}$ is an unnormalized ancilla state.
What we wish to show is that if this entanglement 
introduces no errors into the secret sharing procedure,
then $|\Psi\rangle$ must be a product of a GHZ triplet 
and the ancilla.  This implies that Eve will gain no
information about measurements on the triplet from
observing the ancilla, or, conversely, if Eve is
to gain information about Alice's bit, she must
invariably introduce errors.

First, suppose that Alice, Bob, and Charlie all
measure their particles in the $x$ basis.  If no errors
are to occur we must have that 
\begin{eqnarray}
p(C= +x|A=\pm x,B=\pm x)=1; 
\nonumber \\
 p(C= -x|A=\pm x,B=\mp x)=1,
\label{8}
\end{eqnarray}
where $p(C=+x|A=+x,B=+x)$ is the probability that Charlie
measures $+x$ given that both Alice and Bob measure $+x$,
and the other quantities are similarly defined.  These
equations imply that 
\begin{eqnarray}
P(+x,+x,-x)|\Psi\rangle =0; & \hspace{1cm} &
P(-x,-x,-x)|\Psi\rangle =0; \nonumber \\
P(+x,-x,+x)|\Psi\rangle =0; & \hspace{1cm} &
P(-x,+x,+x)|\Psi\rangle =0 ,
\label{9}
\end{eqnarray}
where $P(+x,+x,-x)$ is the projection onto the subspace
of the three particle-ancilla Hilbert space in which Alice's
particle is in the $+x$ direction, Bob's is in the $+x$
direction, and Charlie's is in the $-x$ direction.  The
other projection operators are defined in a similar manner.
Expressing the conditions in Eq. (\ref{9}) in 
the $z$ basis (the $|0\rangle ,|1\rangle$ basis), we find
that if projection operators corresponding to any of the
vectors
\begin{eqnarray}
\frac{1}{\sqrt{2}}(|000\rangle_{3}-|111\rangle_{3});& 
\hspace{0.5cm}&\frac{1}{\sqrt{2}}(|100\rangle_{3}-
|011\rangle_{3}); \nonumber \\
\frac{1}{\sqrt{2}}(|010\rangle_{3}-|101\rangle_{3});& 
\hspace{0.5cm}&\frac{1}{\sqrt{2}}(|110\rangle_{3}-
|001\rangle_{3}),
\label{10}
\end{eqnarray}
act on $|\Psi\rangle$, the result is zero.

Now suppose that Alice measures her particle in the
$x$ basis, and Bob and Charlie measure theirs in the
$y$ basis.  In order for their to be no errors we
must have that
\begin{eqnarray}
p(C=-y|A=\pm x,B=\pm y)=1;
\nonumber \\
p(C=+y|A=\pm x,B=\mp y)=1,
\label{11}
\end{eqnarray}
which imply that
\begin{eqnarray}
\label{proxy}
P(+x,+y,+y)|\Psi\rangle =0; & \hspace{0.5cm} &
P(-x,-y,+y)|\Psi\rangle =0; \nonumber \\
P(+x,-y,-y)|\Psi\rangle =0, & \hspace{0.5cm} &
P(-x,+y,-y)|\Psi\rangle =0 .
\label{12}
\end{eqnarray}
Again expressing these conditions in the $z$ basis we
find that projection operators corresponding to the
vectors

\begin{eqnarray}
\frac{1}{\sqrt{2}}(|000\rangle_{3}-|111\rangle_{3});& 
\hspace{0.5cm}&\frac{1}{\sqrt{2}}(|100\rangle_{3}-
|011\rangle_{3}); \nonumber \\
\frac{1}{\sqrt{2}}(|010\rangle_{3}+|101\rangle_{3});& 
\hspace{0.5cm}&\frac{1}{\sqrt{2}}(|110\rangle_{3}+
|001\rangle_{3}),
\label{13}
\end{eqnarray}
annihilate $|\Psi\rangle$.

So far we have six vectors to which the three-particle 
part of $|\Psi\rangle$ must be orthogonal.  
A seventh, $(|100\rangle_{3}+|011\rangle_{3})/\sqrt{2}$,
emerges when we demand that no errors occur
when Alice measures her particle in the $y$ direction,
Bob measures his in the $x$ direction, and Charlie
measures his in the $y$ direction.  These conditions
imply that $|\Psi\rangle$ must be of the form
\begin{equation}
|\Psi\rangle = \frac{1}{\sqrt{2}}(|000\rangle_{3} + 
|111\rangle_{3})|R\rangle_{\xi} ,
\label{14}
\end{equation}
i.\ e.\ a product of the GHZ state and an ancilla state,
which is what we wished to show.

Finally, let us conclude this section with a discussion
of the resources necessary to implement quantum secret
sharing protocols.  In order to send a shared key 
containing $N$ bits it is 
necessary to use, on average $2N$ GHZ triplets.  If we 
instead use standard quantum cryptography and the classical 
secret sharing protocol, then either $4N$ entangled pairs, 
using the Ekert procedure [\cite{qcrypt2}], or $4N$
particles, using the BB84 procedure [\cite{qcrypt1}], are
required.  In all cases, the number of particles sent
from Alice to Bob and Charlie is $4N$.  In the GHZ scheme,
once the key has been established, Alice needs to send
$N$ classical bits in order to transmit the message.  These
bits can be sent to either Bob or Charlie using a public
channel.  In the hybrid quantum-classical scheme Alice
must send $2N$ classical bits once keys with Bob and
Charlie have been established - $N$ bits to send the
random string to Charlie and another $N$ bits to send
to Bob the string resulting from the bitwise XOR of 
the message and the random string.  In general, the 
more parts into which the secret is split, the greater 
the difference between the number of classical bits 
which must be sent in the hybrid scheme and in the 
entangled-state scheme ($MN$ versus $N$ for a secret
split into $M$ parts).  
We see that entanglement is
able to act as a substitute for transmitted random bits.

\section{Splitting of Quantum Information}
Now suppose that Alice has a string of qubits she would 
like to send to Bob and Charlie in such a way that they
must cooperate in order to extract the quantum information.
She can use shared GHZ triplets, $|000\rangle_{abc}
+|111\rangle_{abc}$, and a procedure very
similar to quantum teleportation to do this \cite{tele}.  
The no-cloning theorem implies that only one copy of
Alice's qubit can be received, so that either Bob or
Charlie, but not both, will posses the final qubit
\cite{noclon}.  The procedure we shall present is
symmetric in that either party can end up with the
final qubit, but information from the other party is
required before this can happen.  Security could be
enforced by requiring that Bob and Charlie meet in 
person to exchange the final information and put the
qubit to its final use.
Let us now look in detail at the procedure for sending 
one qubit.  We shall first describe the protocol, and 
then discuss the reasons for some of the steps.

Alice begins by taking her qubit, which is in the state
$\alpha |0\rangle_{A}+\beta |1\rangle_{A}$, combining
it with her GHZ particle, and measuring the pair in the
Bell basis
\begin{eqnarray}
|\Psi_{\pm}\rangle_{Aa} & = & \frac{1}{\sqrt{2}}
(|00\rangle_{Aa}\pm |11\rangle_{Aa}); \nonumber \\
|\Phi_{\pm}\rangle_{Aa} & = & \frac{1}{\sqrt{2}}
(|01\rangle_{Aa}\pm |10\rangle_{Aa}) .
\label{15}
\end{eqnarray}
We can determine the effect of this measurement on the
particles which Bob and Charlie possess by expressing 
the entire four-particle state as
\end{multicols}
\vspace{-0.5cm}
\noindent\rule{0.5\textwidth}{0.4pt}\rule{0.4pt}{\baselineskip}
\widetext
\begin{eqnarray}
|\Psi\rangle_{4} & = & \frac{1}{2}[|\Psi_{+}\rangle_{Aa}
(\alpha |00\rangle_{bc}+\beta |11\rangle_{bc})+|\Psi_{-}
\rangle_{Aa}(\alpha |00\rangle_{bc}-\beta |11\rangle_{bc})
\nonumber \\
& &  + |\Phi_{+}\rangle_{Aa} (\beta |00\rangle_{bc}
+\alpha |11\rangle_{bc})+|\Phi_{-}\rangle_{Aa}
(-\beta |00\rangle_{bc}+\alpha |11\rangle_{bc})] .
\label{16}
\end{eqnarray}
\begin{multicols}{2}
At this point Alice does not tell either Bob or Charlie
what the result of her measurement is.  This implies 
that the single-particle density matrixes of both
Bob's and Charlie's particles are $(1/2)I$, where $I$
is the $2\times 2$ identity matrix, so that
at this stage of the procedure neither Bob nor Charlie
has any information about Alice's qubit.  Alice now
tells either Bob or Charlie (she makes the choice at
random) to measure his particle. It is the person
who has not been chosen whose particle will contain 
the final qubit.  The party which 
has been chosen to make the measurement, whom we shall 
assume to be Bob for this particular qubit, now 
measures his particle in the $x$  direction, obtaining
either $|+x\rangle_{b}$ or $|-x\rangle_{b}$.  This
still leaves Charlie's single-particle density matrix
as $(1/2)I$, i.\ e.\  he still has no information
about Alice's qubit.  

In order to reconstruct Alice's qubit Charlie needs two
bits of classical information from Alice (which of the 
four Bell states she found) and one 
from Bob.  Alice first verifies that both parties
have received a particle, which we assume can be done
over a public channel, and then sends Charlie the result
of her measurement.  If Alice's result was either
$|\Psi_{+}\rangle_{Aa}$ or $|\Psi_{-}\rangle_{Aa}$, then
Charlie's single-particle density matrix is
\begin{equation}
\rho_{c}=|\alpha |^{2}|0\rangle_{c}\,_{c}\langle 0|
+|\beta |^{2}|1\rangle_{c}\,_{c}\langle 1| ,
\label{17}
\end{equation}
and if the result was either $|\Phi_{+}\rangle_{Aa}$ or 
$|\Phi_{-}\rangle_{Aa}$, then it is
\begin{equation}
\rho_{c}=|\beta |^{2}|0\rangle_{c}\,_{c}\langle 0|
+|\alpha |^{2}|1\rangle_{c}\,_{c}\langle 1| .
\label{18}
\end{equation}
Charlie now has amplitude information about Alice's
qubit, but knows nothing about its phase.  Bob's
one bit of classical information, in conjunction
with the quantum information he now has, will give him
the phase information and allow him to reconstruct 
Alice's qubit.  In particular, the transformation 
which Charlie should perform in order to obtain
Alice's qubit, up to an overall sign, are
\begin{eqnarray}
 |\Psi_{+}\rangle_{Aa}|+x\rangle_{b}  \rightarrow  & I; 
\hspace{1cm} & |\Phi_{+}\rangle_{Aa}|+x\rangle_{b}
\rightarrow \sigma_{x}; \nonumber \\
|\Psi_{+}\rangle_{Aa}|-x\rangle_{b}  \rightarrow & \sigma_{z}; 
\hspace{1cm} & |\Phi_{+}\rangle_{Aa}
|-x\rangle_{b}\rightarrow \sigma_{x}\sigma_{z};
\nonumber \\
 |\Psi_{-}\rangle_{Aa}|+x\rangle_{b}  \rightarrow & \sigma_{z}; 
\hspace{1cm} & |\Phi_{-}\rangle_{Aa}
|+x\rangle_{b}\rightarrow \sigma_{x}\sigma_{z}; 
\nonumber \\
 |\Psi_{-}\rangle_{Aa}|-x\rangle_{b}  \rightarrow & I; 
\hspace{1cm} & |\Phi_{-}\rangle_{Aa}|-x\rangle_{b}
\rightarrow \sigma_{x} .
\label{19}
\end{eqnarray}
We see, then, that Charlie can reconstruct Alice's 
state but only with the assistance of Bob.  Bob must
both measure his particle and send the result to 
Charlie.  Without Bob's information, Charlie has no
information about the phase of Alice's state.

Let us now discuss this procedure.  We are 
making the assumption that any communication over
a classical channel is insecure.  This means we
cannot consider the simplest method of splitting the 
quantum information in Alice's qubit, which is just
to use standard teleportation with an EPR pair and 
send the classical information to Bob and the 
second particle in the EPR pair to Charlie 
\cite{cleve}.  That is why the procedure we have 
outlined above is somewhat more complicated.  Note
that we could securely implement this protocol if
Alice sent her two bits using standard quantum
cryptography.  She would on average, however, need 
4 particles to do so, and an  entangled pair to 
implement the teleportation procedure.  In addition
this procedure will require that five measurements
be made, on average.  The scheme we have presented
requires a single GHZ triplet, and two measurements.
In effect, it substitutes entanglement for
quantum mechanically implemented classical
communication.

Our next task is to 
see how it protects against cheating and eavesdropping.
Let us first note that is Alice's ability to choose
which particle, Bob's or Charlie's, is to receive
the final qubit prevents
cheating by one of the parties if they manage to get
a hold of both of the particles which Alice sends.  
Suppose, for example, that Charlie is dishonest, that 
he has managed to obtain both particles, and that he 
has sent a particle which he has prepared to Bob.  If
Alice chooses Charlie to receive the qubit, his
cheating will go undetected; one Charlie has the
result of Alice's measurement he has her qubit, and
the result of Bob's measurement is irrelevant.  On
the other hand, if she chooses Bob,
then Charlie has a problem.  At the time he sent the
particle to Bob, Charlie did not know the result of
Alice's measurement, and therefore the particle he
sent to Bob is not in the proper quantum state.  
Alice and Bob can detect this by comparing a subset
of the states Bob received to the ones Alice sent, 
which would reveal Charlie's cheating.

This procedure also guarantees that if an eavesdropper
or a cheater has entangled an ancilla with the 
three-particle state, then errors will be introduced.
If the GHZ state in the above protocol is replaced
by the state in Eq. (\ref{7}), then one can show,
using an argument similar to the one in the previous
section, that if no errors are introduced by the
addition of the ancilla, then the state 
$|\Psi\rangle$ is just a product of the GHZ state and
an ancilla state.  This again implies that measurements
on the ancilla will tell an eavesdropper nothing about
the state of the three particles held by Alice, Bob, 
and Charlie.

\section{Four-Particle GHZ State}
It is possible to generalize this procedure to split
information among more than two people.  Let us look
specifically at the case of three.  Alice starts with 
a four-particle GHZ state,
\begin{equation}
|\psi\rangle_{4}=\frac{1}{\sqrt{2}}(|0000\rangle
+|1111\rangle) ,
\label{20}
\end{equation}
keeps one particle for herself and gives one particle
each to Bob, Charlie, and Diana.  Her object is to
generate a shared key bit which can only be figured
out by Bob, Charlie, and Diana if they cooperate.

A method of accomplishing this can be found by expressing
the state $|\psi\rangle_{4}$ in different combinations of
$x$ and $y$ bases.  If the state is expressed completely
in the $x$ basis, 
\begin{equation}
|\psi\rangle_{4}=\frac{1}{4\sqrt{2}}\left[
\prod_{j}(|+x
\rangle_{j}+|-x\rangle_{j})+\prod_{j}(|+x\rangle_{j}
-|-x\rangle_{j})\right] ,
\label{21}
\end{equation}
where $j$ runs over the set \{Alice, Bob, Charlie,
Diana\}, we see that the right-hand side is an an 
equal superposition of all four-particle basis states,
where each single particle state is in the $x$ basis,
with an even number of $-x$ states.  This means that 
if all four people have each measured their particles 
in the $x$ direction, then Bob, Charlie and Diana
can, by combining their results, determine what
the result of Alice's measurement was.  They simply
count the number of $-x$ measurements, and if it
is even, then Alice must have found $+x$, and if it 
is odd, then Alice must have measured $-x$.  It is
necessary for all three of them to combine their
information in order to determine Alice's result,
no subset will do.  Therefore, Alice has succeeded
in splitting her message into three parts.

In order to foil eavesdroppers and cheaters, the
four parties do not want to use only a single basis,
so we must examine what happens if different 
combinations of $x$ and $y$ bases are used.  
Expressing $|\psi\rangle_{4}$ in the $y$ basis
we find that it is an equal superposition of
all four-particle basis states, where each single
particle state is in the $y$ basis, with an even 
number of $|-y\rangle$ states.  This allows Bob, 
Charlie and Diana to determine Alice's state in 
the same way as in the $x$ basis case.  If two of 
the particles are expressed in the $x$ basis and 
two in the $y$ basis, then we see that 
$|\psi\rangle_{4}$ is an equal superposition of 
the $16$ basis vectors with two particles in the
$x$ basis and two in the $y$ basis (with the same
two in the $x$ basis and the same two in the $y$
basis in each of the four-particle basis vectors)
which have an odd number of minus states.  For 
example, if the first two particles are expressed
in the $x$ basis and the second two in the $y$ basis,
the states $|-x\rangle |+x\rangle |+y\rangle |+y
\rangle$ and $|-x\rangle |+x\rangle |-y\rangle |-y
\rangle$ would appear in the expansion of $|\psi
\rangle_{4}$.  Again, Bob, Charlie, and Diana can
determine Alice's state by counting the number of
minus states which appeared as results of their
measurements.

If three particles are expressed in one basis and
the remaining one in the other, then 
$|\psi\rangle_{4}$ is a superposition of all
$16$ basis vectors.  This means that there are
no correlations among the measurements which will
allow Bob, Charlie, and Diana to infer the result
of Alice's measurement.  If all four parties are
choosing their bases at random, this means that in
half the cases, they will not be able to use the
results.

Summarizing, each of the four parties performs a
measurement on their particle in either the $x$ or
$y$ basis.  They then communicate their choice of
basis to Alice (classically) who decides if the
overall basis choice is a usable one, and she then 
communicates all four basis choices to each of the 
other three parties. Using this information and the
results of their measurements, they can, if they
act in concert, determine the result of Alice's
measurement.  This means that Alice, on the one
hand, and Bob, Charlie, and Diana on the other, 
will have, on repeating this process, a shared 
key.  A calculation similar to the one presented
in Section 2 shows that if an eavesdropper tries
to entangle an ancilla with the four-particle
GHZ state, then she will invariably introduce
errors, and her presence can be detected.

\section{Conclusion}
We have shown that GHZ states can be used to split
information in such a way that if one is in 
possession of all of the parts, the information can
be recovered, but if one has only some of the parts,
it cannot.  This applies to both classical and
quantum information.  In the case of classical
information a shared key can be established between
one party and several others all of whom must work
in concert.  An eavesdropper or a cheater will 
introduce errors and can thereby be detected.  In
the case of quantum information the information in
a qubit is split into two parts so that if the 
parts are recombined, the qubit can be recovered.

This represents a different kind of information 
splitting than occurs in quantum copiers \cite{copy}.
There the object is to split the information in one
qubit into two parts so that each part contains as
much information about the original qubit as 
possible.  However, in that case one cannot 
reconstruct the original qubit by combining the
two copies.

The key point in all of this is that multiparticle
entangled states can be used to split information
into parts.  This can be useful in maintaining
security, as has been shown here, but there may
be applications in the processing of quantum 
information as well.

\acknowledgements
We thank Richard Cleve for helpful and stimulating comments.


\end{multicols}

\end{document}